# A Definition of Vortex Vector and Vortex

Shuling Tian[1,2], Yisheng Gao[2], Xiangrui Dong[2,3], Chaoqun Liu[2*]

[1]*College of Aerospace Engineering, Nanjing University of Aeronautics and Astronautics, Nanjing, Jiangsu, 210016, China*

[2]*Department of Mathematics, University of Texas at Arlington, Arlington, Texas 76019, USA*

[3]*National Key Laboratory of Transient Physics, Nanjing University of Science & Technology, Nanjing, Jiangsu, 210094, China*

Vortex is ubiquitous in nature. However, there is not a consensus on the vortex definition in fluid dynamics. Lack of mathematical definition has caused considerable confusions in visualizing and understanding the coherent vortical structures in turbulence. According to previous study, it is realized that vortex is not the vorticity tube and vorticity should be decomposed into a rotational part which is the vortex vector and a non-rotational part which is the shear. In this paper, several new concepts such as fluid rotation of local point, the direction of fluid rotation axis and the strength of fluid rotation are proposed by investigating the kinematics of fluid element in the 2D and 3D flows. A definition of a new vector quantity called vortex vector is proposed to describe the local fluid rotation. The direction of the vortex vector is defined as the direction of local fluid rotation axis. The velocity components in the plane orthogonal to the vortex vector have zero derivatives along the vortex vector direction. The magnitude of the vortex vector is defined as the rotational part of vorticity in the direction of the vortex vector, which is the twice of the minimum angular velocity of fluid around the point among all azimuth in the plane perpendicular to vortex vector. According to the definition of the vortex vector, vortex is defined as a connected flow region where the magnitude of the vortex vector at each point is larger than zero. The new definition for the vortex vector and vortex follows three principles: 1. Local in quantity, 2. Galilean invariant, 3. Unique. The definitions are carefully checked by DNS and LES examples which clearly show that the new defined vortex vector and vortex can fully represent the complex structures of vortices in turbulence.

## Nomenclature

| | | |
|---|---|---|
| $g_z$ | = | the criterion of rotation of fluid element |
| $g_{z\theta}$ | = | the criterion of rotation of fluid element at an azimuth $\theta$ |
| $Ma$ | = | Mach number |
| $\boldsymbol{P}$ | = | transformation matrix for rotating around Z axis |
| $Q$ | = | second invariant of velocity gradient tensor, a criterion of vortex identification |
| $\boldsymbol{Q}$ | = | transformation matrix |
| $r_x, r_y, r_z$ | = | components of the unit vortex vector |

---

* Corresponding author: Email cliu@uta.edu

| | | |
|---|---|---|
| $\vec{r}$ | = | unit vector of rotation direction in frame $xyz$ |
| $\vec{R}$ | = | rotational part of vorticity, vortex vector |
| $\vec{S}$ | = | non-rotational part of vorticity, shear vector |
| $u, v, w$ | = | components of the velocity in frame $xyz$ |
| $U, V, W$ | = | components of the velocity in frame $XYZ$ |
| $U', V'$ | = | components of velocity in $X'Y'$ plane |
| $\boldsymbol{v}$ | = | velocity in frame $xyz$ |
| $\boldsymbol{V}$ | = | velocity in frame $XYZ$ |
| $x, y, z$ | = | coordinates in the reference frame $xyz$ |
| $X, Y, Z$ | = | coordinates in the reference frame $XYZ$ |
| $X', Y'$ | = | coordinates in the reference frame $X'Y'$ obtained by rotating the $XY$ around $Z$ |
| $\alpha$ | = | local strain rate of fluid element |
| $\beta$ | = | averaged angular velocity of fluid element |
| $\Delta$ | = | vortex-identification discriminant |
| $\Delta t$ | = | small time increment |
| $\varphi$ | = | the angle of phase |
| $\lambda_{ci}$ | = | a criterion of vortex identification, imaginary part of the complex conjugate eigenvalue of velocity gradient tensor |
| $\lambda_2$ | = | a criterion of vortex identification, second-largest eigenvalue. |
| $\lambda_\omega$ | = | "vorticity curvature" criterion vortex identification |
| $\xi_{rot}$ | = | rotational vorticity |
| $\xi_Z$ | = | component of vorticity in Z |
| $\vec{\xi}$ | | vorticity |
| $\omega_\theta$ | = | angular velocity of fluid at azimuth $\theta$ |
| $\omega_{rot}$ | = | rotational angular velocity |
| $\Omega$ | = | a criterion of vortex identification, the ratio of vorticity over the whole motion of fluid element |
| $\nabla \vec{\boldsymbol{v}}$ | = | velocity gradient tensor in the frame $xyz$ |
| $\nabla \vec{\boldsymbol{V}}$ | = | velocity gradient tensor in the frame $XYZ$ |
| $\nabla \vec{\boldsymbol{V}}_\theta$ | = | velocity gradient tensor in the frame rotated from $XYZ$ by an angle $\theta$ |
| $\nabla \times \vec{\boldsymbol{v}}$ | = | the curl of velocity $\vec{\boldsymbol{v}}$, vorticity |
| $|\cdot|$ | = | the magnitude of a vector |

# I. INTRODUCTION

Vortex is a special existence form of the fluid motion featured as rotation of fluid elements and is ubiquitous in the nature. Vortex can be observed in many organized flow structures ranged from hurricanes to tornadoes, from airplane trailing vortices to swirling flows in turbines, and from vortex rings at the exit of a pipe to coherent structures in turbulent boundary layer flow.

Although intuitively a vortex can be easily recognized, it is surprisingly difficult to give an unambiguous definition for vortex[1]. As early as in 1858, Helmholtz proposed the concepts of vortex lines and vortex filaments to investigate the simplest vortex motion.[2] However, so far the problem of the mathematical definition of vortex still remains an open issue in fluid dynamics.[1] The lack of a consensus on the vortex definition has caused considerable confusions in visualizing and understanding the vortical structures, their evolution, and the interaction in complex vortical flows, especially in turbulence.[3,4]

As one of important physical quantities in the fluid dynamics, vorticity is mathematically defined as the curl of velocity vector filed, but has no very clear physical meaning. In classic vorticity dynamics, vorticity is always interpreted as twice the angular velocity of the instantaneous principal axes of the strain-rate tensor of a fluid element. If $\nabla \times \vec{v} = 0$ at every point in a flow, the flow is called irrotational otherwise called rotational, which implies that the fluid elements have a finite angular velocity. Therefore, people always qualitatively identify a vortex as a connected fluid region with relatively high concentration of vorticity and treat vortex as vorticity line or vorticity tube. This may be intuitively reasonable for the large scale vortical structures like tornadoes to equalize vortex and vorticity in early era when people assumed flow is inviscid. However, in 3D and viscous flow vortex cannot be represented by vorticity tube in general. An immediate counter-example is 2D Blasius laminar boundary layer where vorticity is very large near the wall surface, but no flow rotation or vortex is found.[5] In addition, vortices are not necessarily the concentration of vorticity in turbulent flow. Based on DNS results for late flow transition, Wang et al[6] found that vorticity will be reduced when the vorticity lines enter the vortex region and the vorticity magnitude inside vortex is much smaller than the surrounding area, especially near the solid wall. In complex vortical flow, especially in turbulence, the vortical structures cannot be represented by vorticity lines or tubes and the direction of vortex axis is also always different from the direction of vorticity lines or tubes. On the other hand, the existence of a vortex tube does not mean the existence of a vortex, for example, a self-closed ring-like "vorticity tube" exists before the hairpin vortices, but no vortex can be identified by any vortex identification criterion.[6] So, vorticity does not imply the rotation of fluid element and it is not adequate to reveal all types of vortical structures. In general, vortex is a physical phenomenon in nature, but vorticity is a mathematical definition of velocity curl and there is no reason to say vortex can be represented by or equivalent to vorticity tube or vorticity surface.

Researchers have recognized that the vortical structures play an important role in turbulence which has been found to be dominated by spatially coherent and temporally evolving vortical motions, which is popularly called coherent structures. A famous comment is given by Küchemann[7]: "Vortices are the sinews and muscles of turbulence." While in many studies of classic vortex dynamics, many qualitative definitions for vortex are developed, a rigorous quantitative mathematical definition becomes crucial nowadays in understanding vortex structures, their evolution and interactions in turbulence. As Jeong and Hussain[8] stressed, when defining the vortex, the quantitative criteria should satisfy the need for Galilean-Invariance. In recent decades, many vortex identification methods were proposed, mainly including Eulerian method and Lagrangian method[9,10,11]. The $\Delta$-criterion[12-14], $Q$-criterion[15], $\lambda_{ci}$-criterion[16,17], $\lambda_2$-criterion[8], $\Omega$-criterion[18] and the most recent $\lambda_\omega$-criterion[19] are representative among many of the most common Eulerian local vortex identification methods which are based on the local point-wise analysis of the velocity gradient tensor[20]. Based on these methods, a vortex exits in the connected region where the criterion is met. Epps gave a very detailed review of vortex identification methods in Ref. **10** and showed that despite the vast number of criterions have been proposed, the field seems to lack an impartial way to determine which criterion is best. However, though it is hard to judge which method is superior over others, it can be found that all these criterions are described by scalar variable but are not able to give the direction and the exact strength of a vortex which has both direction and strength. Therefore, a precise mathematical definition for vortex is still an open question.

Since the intuitive observation of vortex shows that the vortex has a rotational axis, a natural idea for developing the possible criterion of vortex definition is to find a vector from the decomposition of velocity gradient tensor. As a vector, the vorticity, which is the curl of velocity, is the most straightforward way to be used for the purpose of identifying vortex, but as mentioned above, it cannot distinguish a vortical region from a shear layer, Blasius solution for example. In order to remove shear from the velocity gradient tensor, Kolář[21] proposed a triple decomposition method from which a residual vorticity can be found to represent the pure rigid-body rotation of fluid element. However, the triple decomposition is not unique and a basic reference frame has to be used to solve this problem. Searching for the Euler angles that define the basic reference frame is an expensive optimization problem. In Ref. 22, Kolář et al introduced the concept of average co-rotation of material line segments near a point and applied the average co-rotation to vortex identification. Numerical results[22,23] showed that the method can accurately identify the complex vortical structure, such as hairpin vortex in a turbulent boundary layer flow. However, the averaged co-rotation vector is evaluated by an integral equation and it will be difficult to study the transport of vortex in turbulence by the quantity. Thus, finding a rigorous mathematical definition of vortex is still an open question.

Though vorticity does not always represent rotation of fluid element, Liu et al[18] pointed out that vorticity $\vec{\xi} = \nabla \times \vec{V}$ should be further decomposed into two parts: one is the rotational part $\vec{R}$ contributed to rotation and the other one $\vec{S}$ is non-rotational part (shear) like the vorticity without rotation in laminar boundary layer flows, . However, they didn't show how to decompose the vorticity and did not give formula of $\vec{R}$ and $\vec{S}$. In this paper, based on the physical meaning of vortex, several new concepts such as fluid rotation of local point, the direction of fluid rotation axis and the strength of fluid rotation are proposed by investigating the kinematics of fluid element in 2D and 3D flows. A mathematical definition for vortex vector including the direction and magnitude is presented to describe the fluid rotation in the current paper. The direction of vortex vector is the direction of local fluid rotation axis which is determined by the character that the velocity components in the plane normal to the direction of the vortex vector do not change along the direction of the local fluid rotation axis, in the other words, these velocity components orthogonal to the vortex vector have zero derivatives in the vortex vector direction. This leads to the definition of the vortex vector direction. The magnitude of the vortex vector is measured by the rotational part of vorticity in the direction of the vortex vector. According to the mathematical definition of the vortex vector, vortex is defined as a connected fluid region where the magnitude of the vortex vector at each point is larger than zero.

The paper is organized as follows. The physical meaning of vortex, the method determining the direction of local rotation axis and the strength of fluid rotation are given by the analysis on the motion of the 2D/3D fluid element in Section 2 and the mathematical definition of vortex vector and vortex are given in Section 3. The application of the new mathematical definitions on a number of computational results obtained from DNS and LES are presented in Section 4 and the conclusion remarks are given in last section.

## II. Rotation analysis of fluid element

Intuitively, people recognize vortex as a rotational region, for instance, Lugt[24] stated that a vortex is a multitude of material particles rotating around a common center and proposed the use of closed or spiral streamlines to detect vortices. Though a streamline obviously fails to satisfy requirement of Galilean invariant, it has given an idea that we can define vortex as a connected region where the pseudo-closed streamline around every point can be created by picking this point as the center point to observe the relative motion which can be obtained by local velocity subtracted by the speed of center point. Therefore, the concepts of the single-direction rotation of fluid and the pseudo-closed streamline become a key to define vortex. However, fluid rotation is a phenomenon with an axis, therefore, what the direction of the rotation axis is also an import source for defining vortex. In this section, we will propose how to find the direction of the rotation axis for a 3D fluid element, identify fluid rotation and define the strength of fluid rotation in the plane orthogonal to the rotation axis.

### A. Direction of the rotation axis for a 3D fluid element

The rotation axis of a fluid element is predetermined in a 2D flow which is the normal direction of the 2D plane, but it is not that easy for the 3D flow. Therefore, how to find the rotation axis for a fluid element in a 3D flow is the main objective of this subsection. For a point in a 3D flow field, we can analyze the fluid element centered at the point in any reference frame, but there should be some frames in which it is mostly convenient to carry out the analysis. In followings, we will show how to find the rotation axis for a point by analyzing 3D fluid element in a proper reference frame. Consider a point $P$ in the reference frame $xyz$ and assume that the fluid element centered at $P$ rotates around an axis which is a vector $\vec{r} = r_x\vec{i} + r_y\vec{j} + r_z\vec{k}$ passing through the point P. Here, $\vec{r}$ is defined as the direction of the rotation axis or the direction of the angular velocity at the point $P$, see Fig. 1(a). Since $\vec{r}$ may not be parallel with the axis of a reference frame and it is difficult to analyze the motion of the fluid element at point $P$ in such a frame, we transform the reference frame $xyz$ to a new reference frame $XYZ$ in which the fluid-rotational axis vector $\vec{r}$ is parallel to the axis $Z$, as shown in Fig. 1(b). The sides of fluid element in the new frame are also parallel to the coordinate axis. The velocity gradient tensor $\nabla\vec{v}$ and $\nabla\vec{V}$ in reference frame xyz and XYZ can be written respectively as follows,

$$\nabla\vec{v} = \begin{bmatrix} \dfrac{\partial u}{\partial x} & \dfrac{\partial u}{\partial y} & \dfrac{\partial u}{\partial z} \\[2mm] \dfrac{\partial v}{\partial x} & \dfrac{\partial v}{\partial y} & \dfrac{\partial v}{\partial z} \\[2mm] \dfrac{\partial w}{\partial x} & \dfrac{\partial w}{\partial y} & \dfrac{\partial w}{\partial z} \end{bmatrix}, \qquad \nabla\vec{V} = \begin{bmatrix} \dfrac{\partial U}{\partial X} & \dfrac{\partial U}{\partial Y} & \dfrac{\partial U}{\partial Z} \\[2mm] \dfrac{\partial V}{\partial X} & \dfrac{\partial V}{\partial Y} & \dfrac{\partial V}{\partial Z} \\[2mm] \dfrac{\partial W}{\partial X} & \dfrac{\partial W}{\partial Y} & \dfrac{\partial W}{\partial Z} \end{bmatrix}$$

There is a relation between these two coordinate systems:

$$\nabla\vec{V} = Q\nabla\vec{v}Q^{-1} \tag{1}$$

where $Q$ is the transformation matrix from frame $xyz$ to frame $XYZ$ , and by using Quaternions, we can obtain

$$Q = \begin{bmatrix} \dfrac{r_y^2 + r_z^2 + r_z}{1 + r_z} & -\dfrac{r_x r_y}{1 + r_z} & -r_x \\[3mm] -\dfrac{r_x r_y}{1 + r_z} & \dfrac{r_x^2 + r_z^2 + r_z}{1 + r_z} & -r_y \\[3mm] r_x & r_y & r_z \end{bmatrix} \tag{2}$$

$$Q^{-1} = \begin{bmatrix} \dfrac{r_y^2 + r_z^2 + r_z}{1 + r_z} & -\dfrac{r_x r_y}{1 + r_z} & r_x \\[3mm] -\dfrac{r_x r_y}{1 + r_z} & \dfrac{r_x^2 + r_z^2 + r_z}{1 + r_z} & r_y \\[3mm] -r_x & -r_y & r_z \end{bmatrix} \tag{3}$$

where $\vec{r} = r_x\vec{i} + r_y\vec{j} + r_z\vec{k}$.

In the followings, the motion of the fluid element centered at point $P$ will be analyzed in the reference frame $XYZ$. Consider the fluid element sketched in Fig. 1(b). According to fluid dynamics, the motion of fluid element $ABCD - A'B'C'D'$ depends solely on the velocity gradient tensor $\nabla\vec{V}$, but each term of $\nabla\vec{V}$ is associated with a different motion of the fluid element.

By defining $\vec{r}$ or $Z$ as the unique local rotation axis, there will be no other rotation axes laid down in the X-Y plane which is orthogonal to coordinate axis Z. If $\dfrac{\partial U}{\partial Z} \neq 0$, there must be another rotation axis laid down inside the X-Y plane, which is prohibited.

For example, the sides $AB$, $DC$, $A'B'$, $D'C'$ will have a clockwise rotation in the plane parallel to $YOZ$ plane if $\frac{\partial U}{\partial z} < 0$ as shown in Fig. 1(c). However, since $\vec{R}$ is the unique rotation axis of the fluid element, no other rotation axis is allowed. This implies that if $\vec{r}$ is defined as the direction of the rotation axis at the point $P$, there must be $\frac{\partial U}{\partial z} = 0$. In the same way, we must have $\frac{\partial V}{\partial z} = 0$ if $\vec{r}$ is the direction of the rotation axis at the point $P$, see Fig. 1(d).

As shown in Fig. 1(e) and 1(f), it can be seen that $\frac{\partial W}{\partial x}$ can make the sides $AA'$, $BB'$, $CC'$, $DD'$ have a rotation, but it can't change the direction of the sides $AB$, $DC$, $A'B'$, $D'C'$ and vector $\vec{r}$. There is the same conclusion to $\frac{\partial W}{\partial y}$ and $\frac{\partial W}{\partial y}$ which can't change the direction of vector $\vec{r}$. This implies that $\frac{\partial W}{\partial x}$ and $\frac{\partial W}{\partial y}$ can only generate a shearing in $Z$ direction around the X or Y axis, which does not lead to the change of the direction of vector $\vec{r}$.

Since the component of vorticity in the $Z$ direction can be represented by $\xi_Z = \frac{\partial V}{\partial x} - \frac{\partial U}{\partial y}$, obviously, $\frac{\partial V}{\partial x}$ and $\frac{\partial U}{\partial y}$ can only cause the rotation of the fluid element around the vector $\vec{r}$. The diagonal terms of the velocity gradient tensor, i.e. $\nabla \vec{V}$, $\frac{\partial U}{\partial X}$, $\frac{\partial V}{\partial Y}$, $\frac{\partial W}{\partial Z}$, represent the dilatation of a fluid element, therefore, all of them have no effect on the change of the direction of vector $\vec{r}$.

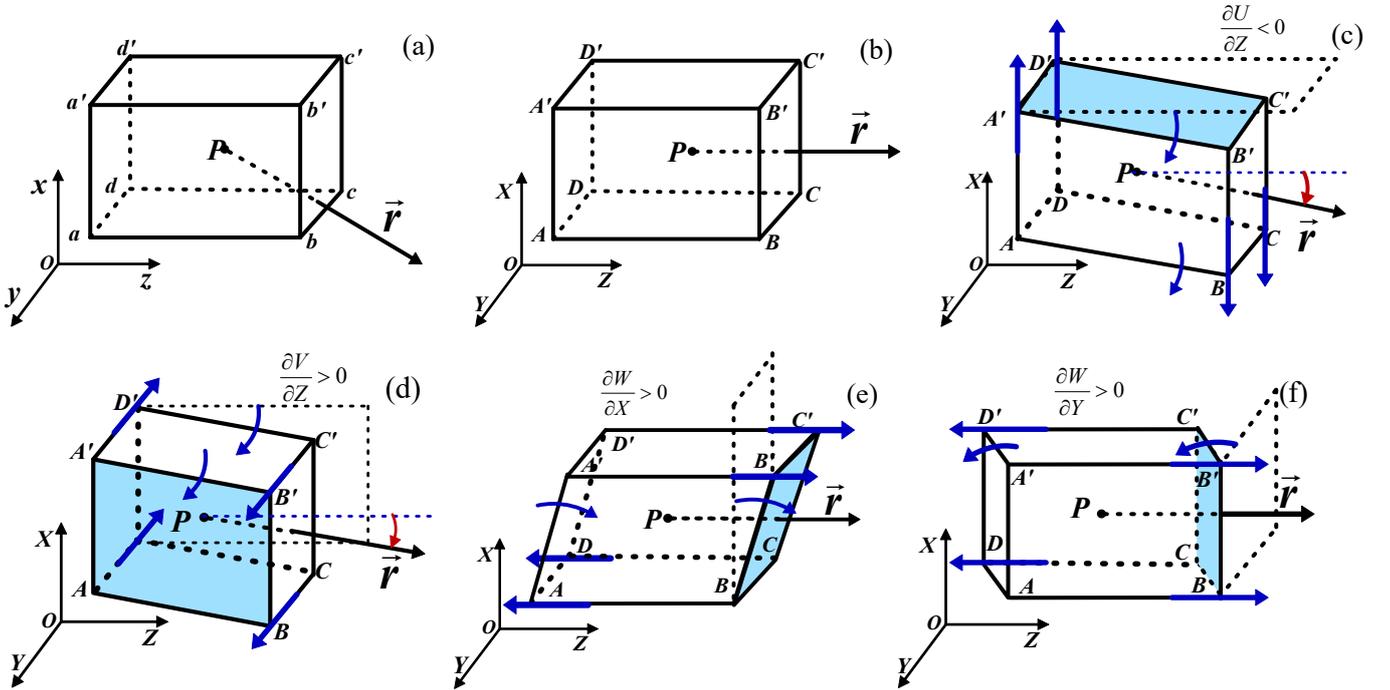

FIG. 1. Rotation and distortion of a fluid element in 3D: (a)fluid element at time t in a general reference frame $xyz$; (b) fluid element at time $t$ in a transformed reference frame $XYZ$; (c)$\frac{\partial U}{\partial Z} \neq 0$ causes a rotation around Y axis to the surface ABCD, A'B'C'D' and to vector $\vec{r}$; (d) $\frac{\partial V}{\partial Z} \neq 0$ causes a rotation around X axis to the surface $ABB'A'$, $DCC'D'$ and vector $\vec{r}$; (e) $\frac{\partial W}{\partial X} \neq 0$ which does not change the direction of vector $\vec{r}$ and only causes shearing in $Z$ direction; (f) $\frac{\partial W}{\partial Y} \neq 0$ which does not also change the direction of vector $\vec{r}$ and only causes shearing in $Z$ direction.

Base on the above analysis, it can be found that in the terms of the velocity gradient tensor $\nabla \vec{V}$, only $\frac{\partial U}{\partial Z}$ and $\frac{\partial V}{\partial Z}$ can cause the change of the direction of vector $\vec{r}$ which is parallel to the $Z$ axis. When $\frac{\partial U}{\partial Z} = 0$ and $\frac{\partial V}{\partial Z} = 0$, $\frac{\partial W}{\partial X}$ and $\frac{\partial W}{\partial Y}$ can only generate shearing in the $Z$ direction and the flow along it is pure shear flow. In this case, there is no rotation around the X and Y axes. So, we can have the following conclusion:

*There must be $\frac{\partial U}{\partial Z} = 0$ and $\frac{\partial V}{\partial Z} = 0$ at point $P$ if the direction of rotation axis at the point is parallel to $Z$ axis.*

Based on the analysis above, we can find the direction of rotation axis $\vec{r} = r_x \vec{i} + r_y \vec{j} + r_z \vec{k}$ for every point by solving the following equations,

$$r_x^2 + r_y^2 + r_z^2 = 1 \qquad (4)$$

$$\frac{\partial U}{\partial Z} = 0 \qquad (5)$$

$$\frac{\partial V}{\partial Z} = 0 \qquad (6)$$

where $\frac{\partial U}{\partial Z}$ and $\frac{\partial V}{\partial Z}$ are determined by Equation 1 in which $\nabla \vec{v}$ is a known tensor and $\frac{\partial U}{\partial Z}$ and $\frac{\partial V}{\partial Z}$ are the functions of the $\vec{r}$. In physics, we can surely find a direction along which the flow is the pure shear flow, therefore, the solution of the set of Equations (4)-(6) must exist. The system of Equations (4)-(6) is a set of nonlinear algebraic equations of vector $\vec{r}$ and it can be solved by using Newton-iterative method. However, the condition that $\frac{\partial U}{\partial Z} = 0$ and $\frac{\partial V}{\partial Z} = 0$ in a reference frame by no means implies that there is a rotation around Z axis but only shows that the fluid has no rotation around X and $Y$ axis. After we find the rotation axis which is a unit vector $\vec{r}$, we should further determine if there is a fluid rotation around the axis aligned to $\vec{r}$ at the point, which can be distinguished by analyzing $\frac{\partial U}{\partial Y}$ and $\frac{\partial V}{\partial X}$ in its normal plane $XY$ of the reference frame $XYZ$. The concept and criterion of fluid rotation will be presented in the next subsection by analyzing the motion of a fluid element in a 2D plane.

**B. Fluid Rotation in 2D plane**

Consider an infinitesimal fluid element moving in two-dimensional flow of the $XY$ plane. At time t, the shape of this fluid element centered at the position $P$ is rectangular and we can assume, without losing generality, that the fluid element is moving upward and to the right, as shown in Fig. 2(a). After a time increment Δt, its position and shape may change, which is determined by the distribution of velocity and its gradient tensor. In the following analysis, we will ignore the translation of the fluid element. Hence, in the $XY$ plane of reference frame $XYZ$, side AB of the rectangular fluid element will have a clockwise rotation if $\frac{\partial U}{\partial Y} > 0$, or have a counter-clockwise rotation if $\frac{\partial U}{\partial Y} < 0$ and have no rotation if $\frac{\partial U}{\partial Y} = 0$. Side AC will have a counter-clockwise rotation if $\frac{\partial V}{\partial X} > 0$, have a clockwise rotation if $\frac{\partial Y}{\partial X} < 0$ and have no rotation if $\frac{\partial V}{\partial X} = 0$. In the classic fluid dynamics, the angular velocity of fluid element is defined as half of vorticity which is $\frac{\partial V}{\partial X} - \frac{\partial U}{\partial Y}$ in 2D. It is easily found that there is an non-zero angular or vorticity when $\frac{\partial V}{\partial X} \neq \frac{\partial U}{\partial Y}$, but the fluid element does not always have a rotation. For instance, it only has a pure shearing in the case of Fig. 2(b) with $\frac{\partial V}{\partial X} = 0$ and irrotational deformation in the case of Fig. 2(c) with $\frac{\partial V}{\partial X} > 0$ and $\frac{\partial U}{\partial Y} > 0$. In this paper, we think the element in the reference frame $XY$ has a rotational motion only when the two sides of AB and AC have a same rotational direction as shown in Fig. 2(d) with $\frac{\partial V}{\partial X} > 0$ and $\frac{\partial U}{\partial Y} < 0$. Generally, we can judge the motion of fluid element in a reference frame by the following criterion parameter,

$$g_Z = \frac{\partial V}{\partial X}\frac{\partial U}{\partial Y}, \tag{7}$$

(i)  Pure shearing or no deformation if $g_Z = 0$;

(ii)  Irrotational deformation if $g_Z > 0$;

(iii)  Rotational deformation if $g_Z < 0$.

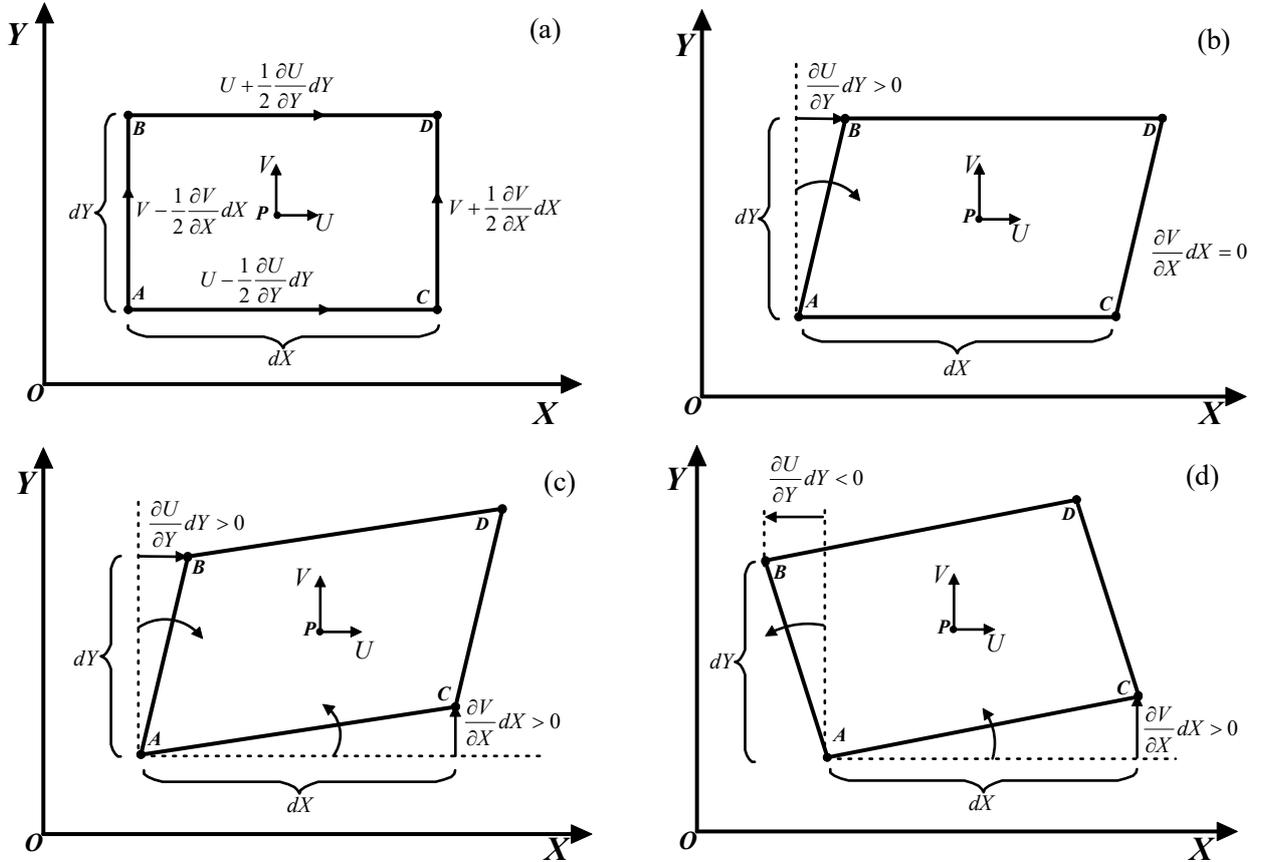

FIG. 2. Rotation and distortion of a fluid element in 2D plane: (a)fluid element at time **t**; (b) pure shearing deformation; (c) irrotational deformation; (3) rotational deformation.

As is known, vorticity $\xi_Z = \frac{\partial V}{\partial X} - \frac{\partial U}{\partial Y}$ in a 2D flow is a Galilean invariant quantity, but $\frac{\partial V}{\partial X}$ or $\frac{\partial U}{\partial Y}$ is not invariant and will change with the rotation of the reference frame. When the frame $XYZ$ is rotated around $Z$ axis by an angle $\theta$, the new velocity gradient tensor is

$$\nabla \vec{V}_\theta = P \nabla \vec{V} P^{-1}, \tag{8}$$

where $P$ is the transformation matrix around $Z$ axis and there are

$$P = \begin{bmatrix} cos\theta & sin\theta & 0 \\ -sin\theta & cos\theta & 0 \\ 0 & 0 & 1 \end{bmatrix}, \qquad P^{-1} = \begin{bmatrix} cos\theta & -sin\theta & 0 \\ sin\theta & cos\theta & 0 \\ 0 & 0 & 1 \end{bmatrix}.$$

So, we can obtain

$$\frac{\partial V}{\partial X}\big|_\theta = \alpha\, si\, n(2\theta + \varphi) + \beta \tag{9}$$

$$\frac{\partial U}{\partial Y}\big|_\theta = \alpha\, si\, n(2\theta + \varphi) - \beta \tag{10}$$

where

$$\alpha = \frac{1}{2}\sqrt{\left(\frac{\partial V}{\partial Y} - \frac{\partial U}{\partial X}\right)^2 + \left(\frac{\partial V}{\partial X} + \frac{\partial U}{\partial Y}\right)^2}, \tag{11}$$

$$\beta = \frac{1}{2}\left(\frac{\partial V}{\partial X} - \frac{\partial U}{\partial Y}\right), \tag{12}$$

and

$$\tan\varphi = \frac{\frac{\partial V}{\partial X} + \frac{\partial U}{\partial Y}}{\frac{\partial V}{\partial Y} - \frac{\partial U}{\partial X}}. \tag{13}$$

Based on Equations (9) and (10), the criterion of rotation of fluid element in Equation (7) becomes

$$g_{z\theta} = \frac{\partial V}{\partial X}\big|_\theta \frac{\partial U}{\partial Y}\big|_\theta = \alpha^2 \sin^2(2\theta + \varphi) - \beta^2. \tag{14}$$

It can be seen that the sign of $g_{z\theta}$ will change with $\theta$ if $\alpha^2 - \beta^2 > 0$. This implies that at the same point, the fluid elements at the different azimuth may have different types of motion as we look at the fluid motion in different reference frames, which is then not reference frame invariant. For example, as shown in Fig. 3(a), for the same point $P$, we can consider two different fluid elements in two different frames. After a time increment $\Delta t$, the fluid element in $XY$ has a rotational deformation as $g_z < 0$, see black element in Fig. 3(b), however, the fluid element centered at the same point but in the frame rotated by an angle $\theta$ has an irrotational deformation as $g_z > 0$, see blue element in Fig. 3(b). It reveals that the motion character of fluid element centered at a point is related to the azimuth $\theta$, or, in other words, the motion is not invariant in different reference frames. However, if $\alpha^2 - \beta^2 < 0$, $g_{Z\theta} < 0$ for any $\theta \in [0,2\pi)$, which means that the fluid element at any azimuth $\theta$ has a single rotation direction. In this paper, we consider that a point is fluid-rotational if the rotation direction is invariant in any reference frames. Therefore, based on the distribution of the criterion $g_{Z\theta}$, we propose a definition to identify if a point is fluid rotational in a 2D plane as follows,

**Definition 1:** *a point is **fluid-rotational** in a 2D plane if the velocity gradient tensor at the point meets*

$$g_{Zmax} = \left(\frac{\partial V}{\partial X}\big|_\theta \frac{\partial U}{\partial Y}\big|_\theta\right)_{max} = \alpha^2 - \beta^2 < 0, \tag{15}$$

*where $\alpha$ and $\beta$ are defined in Equation (11) and (12).*

The definition implies that if a point is fluid-rotational, in any 2D reference frames, the fluid element has a rotational deformation and the rotational direction is invariant. In other words, the fluid at any azimuth $\theta$ around the point has the same rotation direction relative to the point and the single direction pseudo-closed streamline can be formed by using the velocity component in the tangential direction relative to the fluid-rotational point. If we do not analyze the rotation character of a fluid rotational point by using fluid element, $\frac{\partial V}{\partial X}$ can be regarded as the fluid's angular velocity relative to the point at azimuth $\theta$, $\omega_\theta$, so we have

$$\omega_\theta = \frac{\partial V}{\partial X}\big|_\theta = \alpha sin(2\theta + \varphi) + \beta \tag{16}$$

It can be easily obtained from Equation (16) that the direction of angular velocity is invariant for the fluid-rotational point but it is not for non-fluid-rotational point. For instance, there are two points in the frame $XYZ$ and the velocity gradient tensors at the points are

$$\nabla \vec{V}|_1 = \begin{bmatrix} 0.08 & -0.3 & 0 \\ 0.16 & -0.05 & 0 \\ 0 & 0 & 0 \end{bmatrix}, \qquad \nabla \vec{V}|_2 = \begin{bmatrix} -0.25 & -0.025 & 0 \\ 0.125 & 0.150 & 0 \\ 0 & 0 & 0 \end{bmatrix}.$$

Obviously, there are $\frac{\partial U}{\partial Z} = 0$ *and* $\frac{\partial V}{\partial Z} = 0$ for the two points and the direction of rotation axis is only along $Z$ axis if they are fluid-rotational. We can identify their rotational character by analyzing their angular velocity. The distributions of the angular velocity and relative velocity component in the tangential direction around the points are shown in Fig. 4 and it can be seen that the fluid has an invariant direction rotation (clockwise or counter-clockwise) around the first point and (see Fig. 4(a) and (b)), but the rotation direction is variant with the azimuth $\theta$ around the second point (see Fig. 4(a) and (b)). And there are $\alpha^2 - \beta^2 < 0$ for the first point and $\alpha^2 - \beta^2 > 0$ for the second point. By the definition, we can judge the first point in Fig. 4 is a fluid-rotational one, but the second point is not.

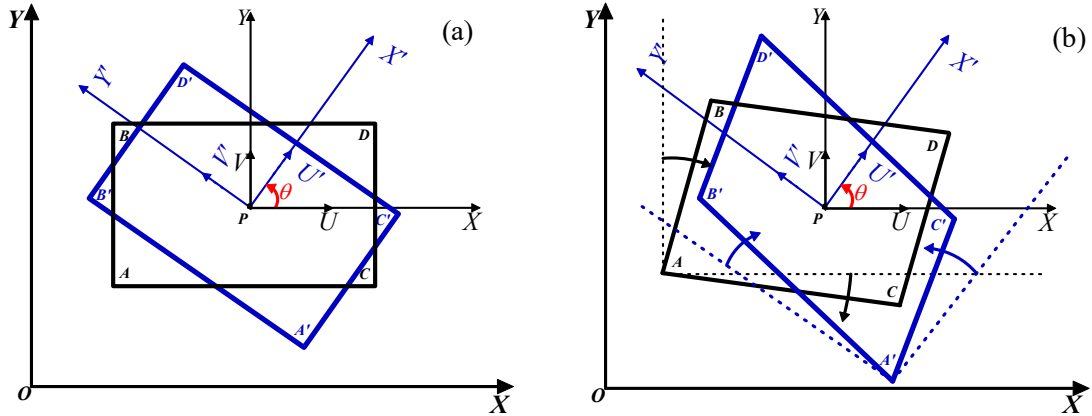

FIG. 3. Rotation character of a fluid element in two different reference frames: (a) rectangular fluid element at time $t$; (b) rotation and distortion after a time increment $\Delta t$.

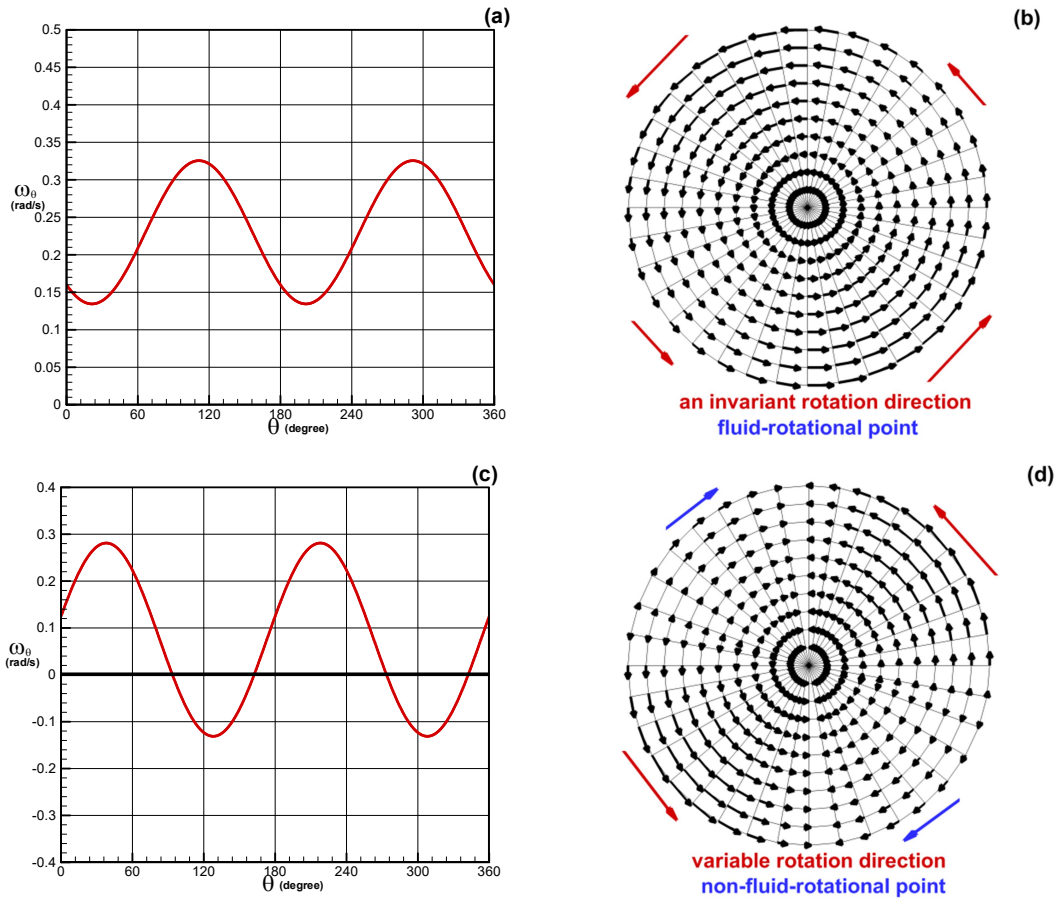

FIG. 4. Rotation character of fluid around two points: (a) angular velocity distribution of fluid around the first point; (b) distribution of velocity components in tangential direction around the first point; (c) angular velocity distribution of fluid around the second point; (b) distribution of velocity components in tangential direction around the second point.

Different from the rotation of rigid body which has the same angular velocity for every point, the angular velocity of fluid, even if at the same point, changes with azimuth and it is not able to directly determine the fluid-rotational strength for a point. How to determine the strength of fluid rotation is another issue. Based on the idea of Liu et al[18], for the fluid-rotational point, we use the minimum angular velocity of Equation (16) as the fluid-rotational angular velocity, so we give the following definition:

**Definition 2**: *fluid-rotational angular velocity* $\omega_{rot}$ *for a fluid-rotational point in the 2D plane is defined as*

$$\omega_{rot} = \begin{cases} \beta - \alpha, & if \ \beta > 0 \\ \beta + \alpha, & if \ \beta < 0 \end{cases}, \tag{17}$$

*where $\alpha$ and $\beta$ are defined in Equation (3) and (4) and they meet Equation(15).*

Based on Definition 1 and 2, the fluid-rotational angular velocity is zero for the non-fluid-rotation point ($\alpha^2 - \beta^2 \geq 0$). For the 2D flow, the direction of the fluid-rotational angular velocity of a fluid-rotational point is perpendicular to the 2D plane, and so we can have the following conclusions,

(i)   if $\omega_{rot} > 0$, the fluid-rotation is counter-clockwise;

(ii)  if $\omega_{rot} < 0$, the fluid-rotation is clockwise;

(iii) if $\omega_{rot} = 0$, the point  is not fluid-rotational.

The magnitude of vortex, or the rotational vorticity, can be defined as follows,

**Definition 3**: *rotational vorticity* $\xi_{rot}$ *for a point in the 2D plane is defined as the twice of fluid-rotational angular velocity* $\omega_{vort}$, *here*

$$\xi_{rot} = 2\omega_{rot}, \tag{18}$$

By definition 3,  $\xi_{ro}$  is the rotational part of vorticity and it has the same form as  the residual vorticity proposed by Kolář et al in Ref. **22** where they think vorticity dominates over strain rate for the co-rotation case ($\beta^2 > \alpha^2$). However, our definition is based on the invariant feature of the fluid element rotation direction and critical value in different rotating reference frames to determine the fluid rotation and strength of the vortex (or rotational vorticity part) which is different from residual vorticity concept and the 3D rotational axis definition is completely different from the triple tensor decomposition.

After moving the rotational part of vorticity, the left part of vorticity represents pure shear of flow in the plane. By using the direction of the fluid rotation axis and the rotational vorticity, we can define the new vector quantity which can describe the rotation of fluid, which will be shown in the next section.

## III. Definition of vortex vector and vortex

### A.  Definition of vortex vector

In Section 2, by analyzing the physical meaning of fluid element motion in 3D and 2D, we present several concepts of rotation of fluid, such as fluid rotation, the direction of fluid rotation axis, angular velocity and rotational vorticity. By these concepts and definitions, we will propose a definition of a new vector quantity which is called vortex vector and will be used to definite vortex in this paper.

**Definition 4**: *a vortex vector $\vec{R}$ at a point is a local vector. The direction of the vortex vector is the potential direction of fluid rotation axis $\vec{r}$. The magnitude of vortex vector is the rotational vorticity $\xi_{rot}$.*

By the definition above, we can find the new defined vortex vector by the following steps and methods:

**(1) Determine the direction of vortex vector;**

The direction of vortex vector is determined by the potential direction of fluid rotation axis $\vec{r}$. It can be obtained by solving the system of equations (4)-(6). From the analysis of Section 2, it can be seen that there may not be fluid-rotational even if a unit vector $\vec{r}$ which satisfies equations (4)-(6) is found and we should further identify the fluid-rotation by the following steps.

**(2) Identify fluid-rotation;**

After obtaining the potential direction of fluid rotation axis $\vec{r}$, transform the reference frame *xyz* to the frame *XYZ* in which $\vec{r}$ is parallel to *Z* axis by Equation(1)-(3). We can identify the fluid-rotation at a point in the XOY plane by criterion of the Definition 1 and the Equation (15).

**(3) Determine magnitude of vortex vector.**

The magnitude of the vortex vector is defined as the rotational vorticity which is determined by the critical value of the rotational angle speed and is unique and invariant.

By the three steps, we can obtain the vortex vector $\vec{R}$ by the following equation,

$$\vec{R} = \xi_{rot}\vec{r}, \tag{19}$$

For the fluid-rotational point, the vortex vector $\vec{R}$ can be determined by Equations (17)~(19), however the vortex vector is zero for the non-fluid-rotational point which does not satisfy Equation (15). As we addressed in Section 2 that the solution of the set of Equations (1), (4), (5) and (6) must exist at any point and the solution is also unique for the fluid rotational point. If there are two different directions of rotational axes, e.g. $\vec{r}_1$ and $\vec{r}_2$, with the vortex vectors $\vec{R}_1 \neq 0$ and $\vec{R}_2 \neq 0$ respectively, there must exist a component of vortex vector $\vec{R}_1$ in the normal plane of $\vec{r}_2$ and thus the velocity in the plane must change in the direction of $\vec{r}_2$ or its derivative along $\vec{r}_2$ is not zero. This means that $\vec{r}_2$ is not the direction of the fluid rotational axis, which contradicts the hypothesis that $\vec{r}_2$ is the direction of the fluid rotational axis. So, for the fluid rotational point, the solution of the set of Equations given above must be unique.

After finding the rotational part of vorticity, we can obtain the non-rotational part of vorticity $\vec{S}$ which is pure shear of fluid by

$$\vec{S} = \nabla \times \vec{v} - \vec{R} \tag{20}$$

## B. Definition of vortex

In the above subsection, we give a new vector quantity, *vortex vector*, whose direction is that of fluid-rotation axis and whose magnitude is the strength of fluid rotation. Vortex vector can be used to define vortex which is the building bricks of the turbulence coherent structure, as follows

**Definition 5**: *A vortex is a connected region where* $\vec{R} \neq 0$.

After having the definition of vortex, how to visualize the vortices in the flow field is also a problem. Similar to vorticity line and vorticity tube, vortex line and vortex tube can be used to display the vortical structures and they are defined as follows,

A *vortex line* is a curve whose tangent at any point is in the direction of the vortex vector at that point. The concept of vortex tube is related to vortex line. Consider an arbitrary closed curve in three-dimensional space, the vortex lines which pass through all points on the curve form a tube in space and such a tube is called a *vortex tube*.

Since we have defined vortex vector, we have two different ways to describe the vortex structure in a flow field: using vortex lines and tubes or using the iso-surface of the magnitude of vortex vector which represents the strength of fluid-rotation. In the next section we will demonstrate the capability of vortex vector in describing the complex vortex structure in different flow fields. Mainly, we will use the iso-surface of the vortex strength while leave the vortex lines and tubes for future applications.

## IV. Applications of the definition of vortex vector

As a vector quantity, vortex vector represents the local rotation of fluid. Like most of vortex identification methods, the iso-surface of the magnitude of vortex vector can show the core of vortex. In this section, the definition of vortex vector presented in section II and III will be evaluated and compared by using 2D/3D DNS /LES databases which have been tested in Liu's previous work by different vortex identification methods[18,25].

### A. 2D Blasius-profile temporal mixing layer

As the first example, we consider a 2D mirrored Blasius-profile shear layer excited with a perturbation to induce the rollup of shear layer and form vortices. The unsteady 2D flow field with an inflow *Ma*=0.5 is numerically simulated by the code "DNSUTA" in which NS equations are solved using a sixth-order compact scheme in the spatial discretization and a third order TVD Runge–Kutta scheme in the tempral discretization. A local refined grid is generated with a dimension $256 \times 512$ in the streamwise and normal directions respectively.

The evolution of the flow visualized by vorticity contours is shown in Fig. 5 for four time instants. At the initial stage, there is plenty of voritcity in the shear layer, but the distribution of vortex vector $\boldsymbol{R}$ does not show any vortex existence in the shear layer (see Fig. 5(a)). This clearly implies that the existence of vorticity does not mean the existence of vortex and the vorticity can't fully describe the flow phenomenon of vortex. In the following instants, there are two clockwise vortices which are shown as the black closed lines in the Fig. 5(b)~(d) and the strength of the vortices become stronger with time.

Consider the instant of flow field as shown in Fig. 3(c), the vortices are identified by $Q$-criterion, $\Omega$-criterion and vortex vector $\vec{R}$, respectively. The distributions of $Q$, $\Omega$ and $\vec{R}$ are shown in Fig. 6. In the Fig. 6(a)~(c), the black lines are the boundary of vortices identified by $Q = 0.0$, $\Omega = 0.5$ and $|\vec{R}| = 0.0$, respectively. It can be seen that all those methods can give the same vortical structures in the 2D low speed flow. In fact, we can obtain the same conclusions by substituting the continue equation of 2D incompressible flow into Equation (9) which will become Q-criterion. Note that $|\vec{R}|$ and Q will not be same for 3D flow or compressible flow. Fig. 6(d) shows the distribution of the non-rotational part of vorticity which describes the shearing of flow. It can be seen that the distribution of the non-rotational part of vorticity is very similar to that of vorticity. This reveals that the fluid rotation is different from the rotation of rigid body and there is a strong shearing with the rotation of fluids in the vortex.

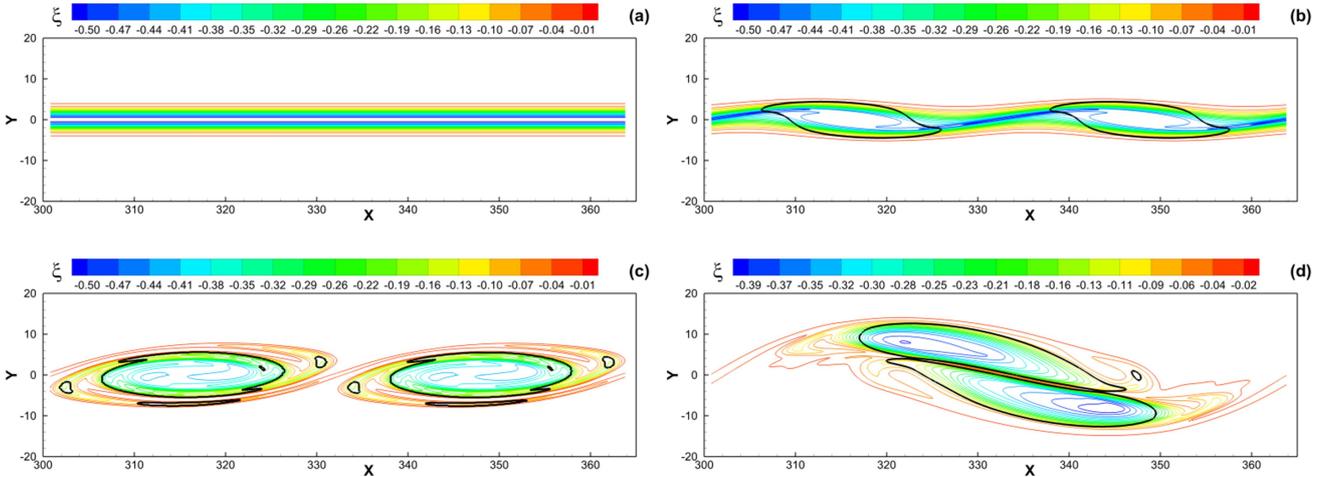

FIG. 5. The temporal evolution of shear layer: (a) the distributions of vorticity at initial stage; (b) vorticity distribution with weak vortex generated during shear layer rolling up; (c) contours of vorticity with a pair of vortices generated; (d) contours of vorticity in the paring of the two vortices.

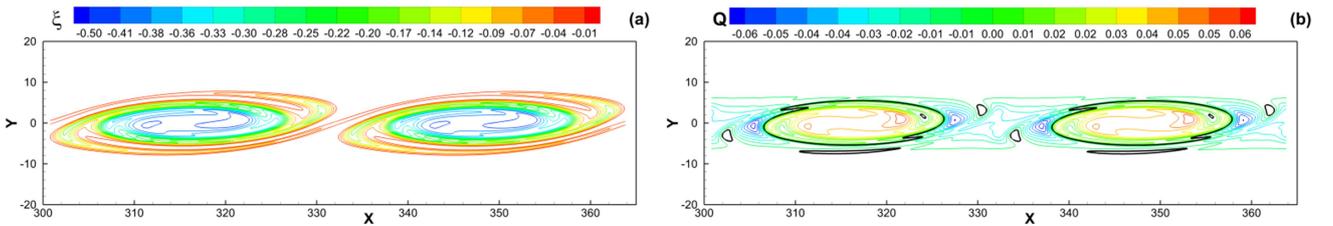

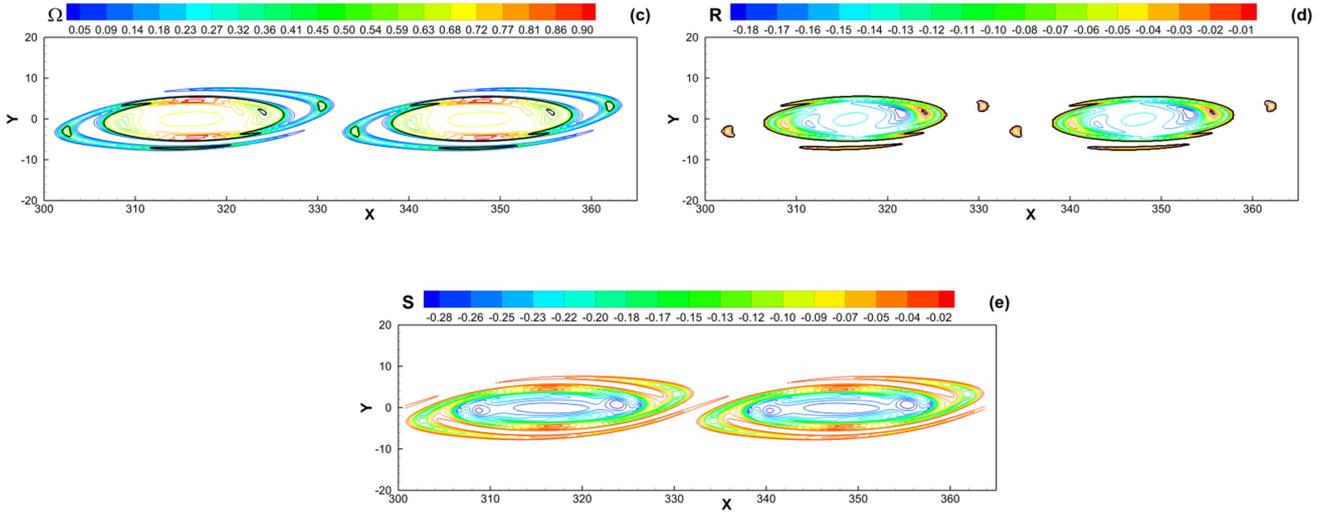

FIG. 6. Distributions of a pair of vortices: (a) contours of vorticity which is not able to tell where the vortices are; (b) contours of Q-criterion with the boundary of vortex as the black lines which are the iso-lines of Q = 0.0; (c) contours of Ω-criterion with the boundary of vortex as the black lines which are the iso-lines of Ω = 0.5; (d) contous of vortex vector with a natural vortex boundaries which are the iso-lines of $|\vec{R}| = 0.0$; (e) contours of the non-rotational part of vorticity.

As we present in Section 2, if a point is fluid rotational, the fluid flow in a neighborhood of the point has a tendency to move around it or we can generate pseudo-closed streamlines based on the velocity relative to the point. At the instant of Fig. 5(c), we consider two points A and B as shown in Fig. 7. Point A is out of the regions of vortices and point B lies in the left vortex, see Fig. 7(a). Using the velocity relative to each point respectively, the streamlines are created near point A (see Fig. 7(b)) and point B (see Fig. 7(c)). As we can see, there is no pseudo-closed streamline around point A, but the pseudo-closed streamline can be formed around point B, which also shows that point B is fluid rotational and point A is not.

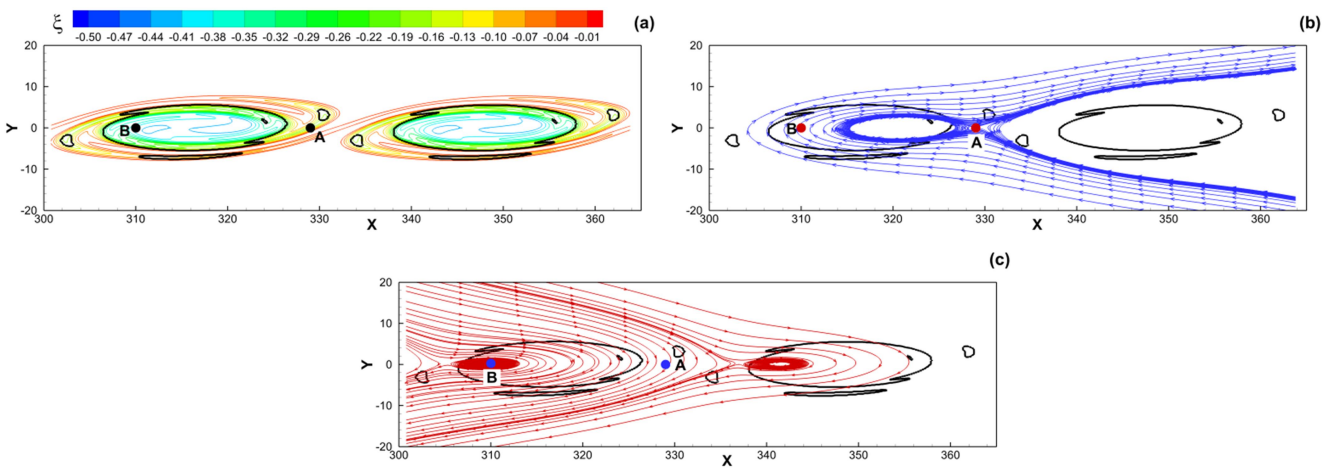

FIG. 7. Pseudo-streamlines based on the relative velocity: (a) the positions of points A which is out of the vortex and B which is in the vortex; (b) no closed pseudo-streamline around non-fluid rotational point A; (c) closed pseudo-streamline around the fluid rotational point B.

From the example above, it can be seen that the definition of vortex vector can precisely describe vortex and its magnitude represents the strength of vortex. In the vortex, each point is fluid rotational, which means that the fluid in a neighborhood of that point has the same direction of tangential velocity relative to that point. With the newly defined vector quantity, vortex vector can be used to quantitatively study vortices.

## B. vortices in late boundary layer flow transition

For the 2D vortical structures, the direction of the vortex vector is predetermined, which is vortical to the flow plane. However, in the 3D flow, the directions of vortex cores are arbitrary and they may change in time and space. By the definition of vortex vector, how to determine the vortex direction is a key issue. As the second example, we consider the vortices in a late boundary layer transition which is a process of vortex "buildup". In the process of boundary layer transition, Λ-shaped vortex and hairpin vortex can be commonly observed not only in the experiment but also in the DNS simulation and they will be used to show the ability of the newly defined vortex vector in representing the complex vortex structure. In the following, the DNS data[25] of late boundary layer flow transition obtained by DNSUTA is applied to provide the structures of vortices.

As one of the most popular vortex identification methods, Q-criterion is selected to compare the ability of capturing the vortical structures in turbulence with our newly defined vortex vector. In fluid dynamics, the iso-surface of Q is always used to illustrate the cores of vortex. The typical structure of Λ vortex is illustrated in Fig. 8 where the left blue surface is the iso-surface of $Q = 0.02$ and the right green surface is the iso-surface of the magnitude of vortex vector $|\vec{R}| = 0.075$. Both of them can illustrate the similar structure of vortex cores. However, being different from the Q-criterion vortex identification method, the magnitude of vortex vector represents the strength of fluid rotation and the strength is same on the iso-surface of $|\vec{R}|$. The larger head of vortex vector iso-surface reveals that there is a stronger rotation than in the tail, which the traditional vortex identification method can't capture.

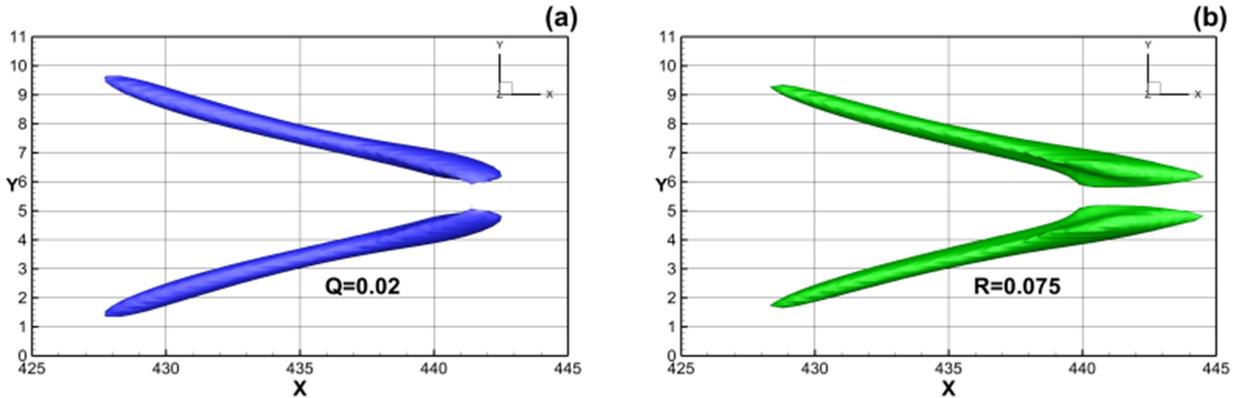

FIG. 8. Typical structure of Λ vortex visualized by (a) iso-surface of $Q = 0.02$; (b) iso-surface of the magnitude of vortex vector $|\vec{R}| = 0.075$.

In classic vorticity dynamics, the vortices are always regarded as vorticity lines or tubes. However, they can't fully describe the real vortex structures, even if as simple as a Λ vortex. As shown in Fig. 9(a), the vorticity lines are created through points in the core of Λ vortex and it can be seen that the structure of the vorticity lines is totally different from that of Λ vortex. There are only a small section of vorticity lines roughly aligned with the cores of vortex, but the new vortex lines defined in Section 3 of this paper have the same structure as the cores of Λ vortex, see Fig. 9(b) in which the vortex lines are created inside the vortex core, which are all aligned with the vortex core direction. In Fig. 9, the vorticity lines and vortex lines are colored with the magnitude of vorticity and vortex vector, respectively. It can be seen that as the rotational part of vorticity, i.e. vortex vector, can not only fully describe the structure of vortex but also quantitatively give the strength of fluid rotation where the red represents stronger rotation.

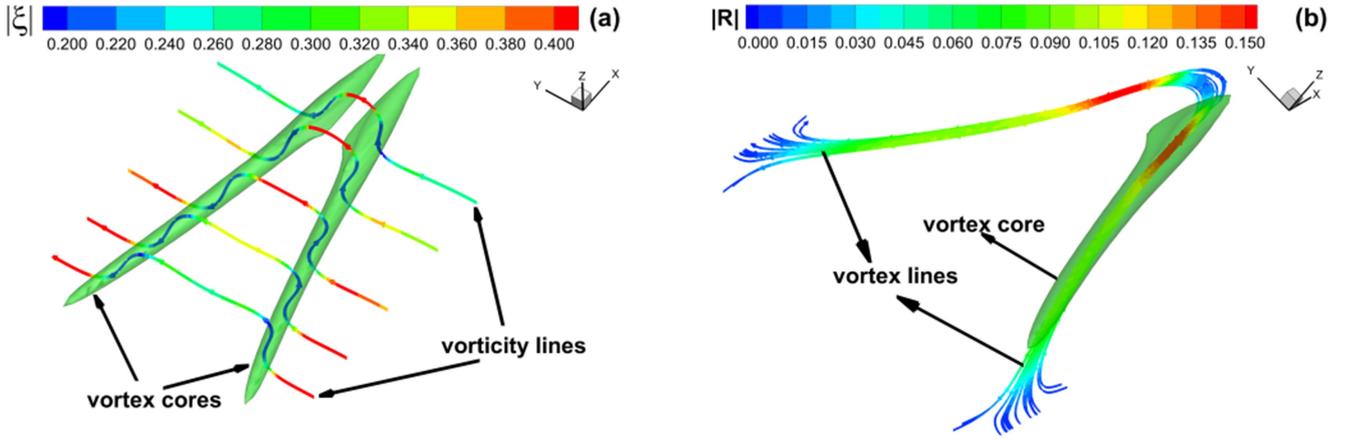

FIG. 9. Compare vorticity lines with vortex lines: (a) vorticity lines which are created through points in the vortex core are not aligned with the cores of Λ vortex legs and can't be used to represent vortex; (b) vortex lines which are also created through the points in the vortex cores are aligned with the cores of Λ vortex legs and can represent the structure of vortex.

Different from the vortex identification methods, such as Δ-criterion, Q-criterion, $\lambda_{ci}$-criterion, $\lambda_2$-criterion and Ω-criterion, which are scalar quantity, vortex vector, as a vector quantity, can also show the direction of local fluid rotation. In order to further compare the difference between vorticity and vortex vector, the vectors on the vortex cores which are illustrated by the iso-surface of $|\vec{R}| = 0.075$ are shown in Fig. 10. As we can see, the vector of vorticity is not aligned (even orthogonal to) with the vortex core (see Fig. 10(a)), but the vortex vector is highly aligned with the all parts of Λ vortex cores (see Fig. 10(b)). The shear vector is also shown on the vortex cores and it has a very similar distribution as vorticity, see Fig. 10(a) and (c). When we put the vectors of vorticity $\nabla \times \vec{v}$, vortex vector $\vec{R}$ and shear vector $\vec{S}$ at the same point and create the vortex line and vorticity line through that point, it can be seen clearly that vortex line and vortex vector are aligned to the vortex core, vortex vector can represent fluid rotation and the vortex vector is the rotational part of vorticity. However, the vector of vorticity has a different direction from the vortex and it further shows that vorticity can't represent rotation of fluid. On the iso-surface of $|\vec{R}| = 0.075$, shear vector is the main part of vorticiy, as shown in Fig. 10(d). The vector distributions of vortex vector and vorticity further imply that vortex vector can fully describe vortex structure, but vorticity cannot for 3D vortices. Actually, we have the formula: $\nabla \times \vec{v} = \vec{S} + \vec{R}$, which means vorticity should be decomposed into vortex vector and shear.

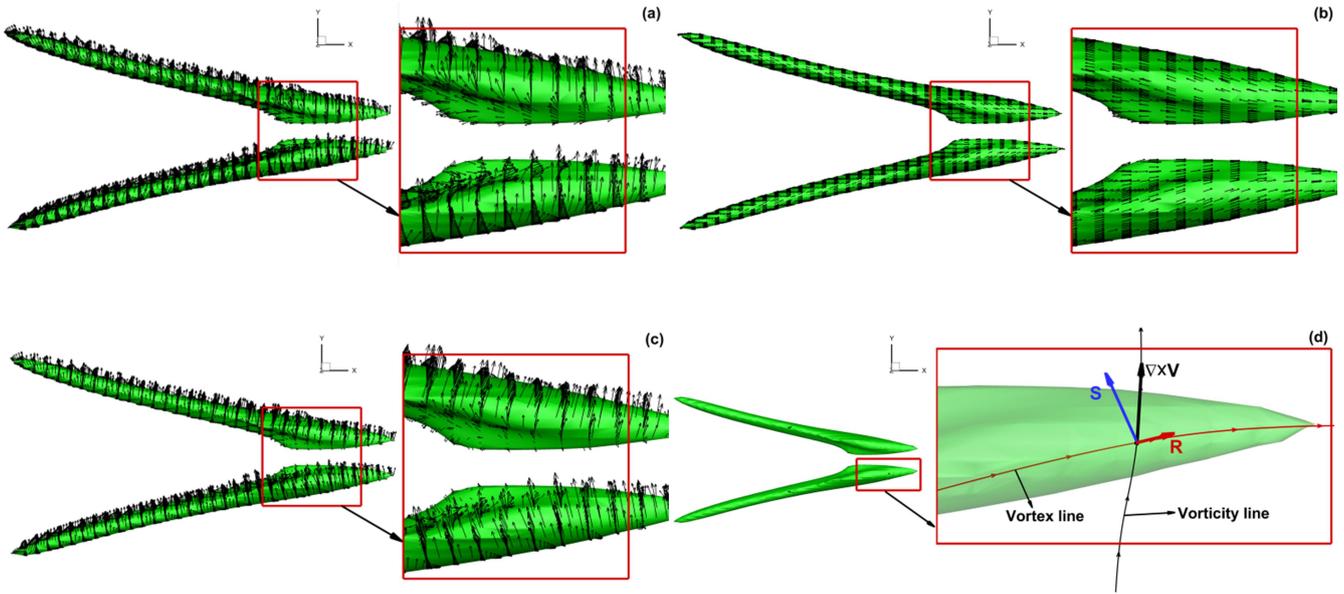

FIG. 10. Compare the vector of vorticity with the vortex vector on the cores of vortex which are illustrated by the iso-surface of $|\vec{R}| = 0.075$: (a) the vector of vorticity which is not aligned with (even orthogonal to) the vortex core and can't fully describe vortex; (b) the vector of vortex vector which is aligned to the core of $\Lambda$ vortex can represent vortex;(c)the vector of shear vector which is the non-rotational part of vorticity; (d)compare the vectors of vorticity, vortex vector and shear vector at a point.

In the process of late boundary layer transition, hairpin vortex is a common vortex structure[26] and the flow structure is dominated by the hairpin vortices after the overshoot of skin-friction[27]. In the natural flow transition, the hairpin vortices always appear in packets, as shown in Fig. 11 where a typical packet of hairpin vortices is visualized by using the iso-surface of $Q = 0.02$ (see Fig.11 (a)) and the magnitude of vortex vector $|\vec{R}| = 0.075$ (see Fig.11 (b)). Similar to the example of $\Lambda$ vortex, both of methods can give a similar structure of vortices. However, the iso-surface of vortex vector can show more small vortices while keeping the same shape and size of large vortices than the Q-criterion. This implies that the iso-surface of the magnitude of vortex vector can correctly and accurately represent the topology and geometry of the vortex core structures of the complex vortices.

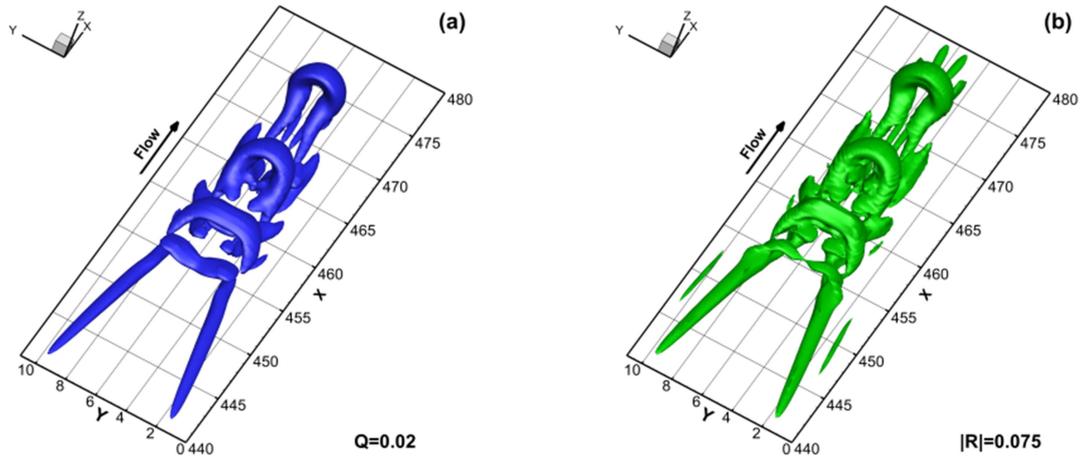

FIG. 11. Typical structure of a packet of hairpin vortices illustrated by (a) iso-surface of Q = 0.02; (b) iso-surface of the magnitude of vortex vector $|\vec{\boldsymbol{R}}| = 0.075$.

Fig. 12 shows the structures of the vorticity lines and vortex lines which are all created through the same points. As can be seen, both vorticity lines and vortex lines can represent the topology of the ring of the hairpin vortex, but only vortex lines are consistent with all vortex cores. The analysis on the distribution of vorticity and vortex vectors shows that the ring of hairpin vortex is a very strong vortex core and most part of vorticity is rotation vorticity or vortex vector. However, the other parts of the packet of hairpin vortex are weak and the vortex vector is only the small part of vorticity, so that the vorticity lines through those points in the weak vortex cores are not aligned to the vortex cores and only vortex lines are always consistent with the vortex cores there. It can be further verified by the vector distribution of vortex vector, see Fig. 13, which illustrates vortex vectors on the surface of the core of the hairpin vortex packet.

To show the ability of vortex vector in describing the complex structure of vortices, the vortical structures at one instant in the "buildup" of the laminar-turbulence flow transition are visualized in Fig.14 where the vortex cores are illustrated as the iso-surface of the magnitude of vortex vector $|\vec{\boldsymbol{R}}| = 0.075$. It can be found that the structures of complex vortex can be correctly captured by the vortex vector.

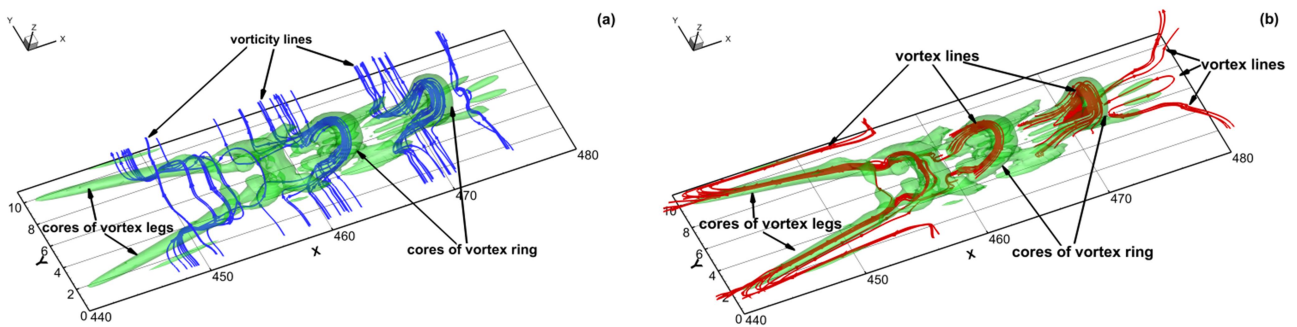

FIG. 12. The distribution of vorticity lines and vortex lines: (a) vorticity lines which are not consistent with the cores of vortex; (b) vortex lines which are aligned to the cores of the complex vortices.

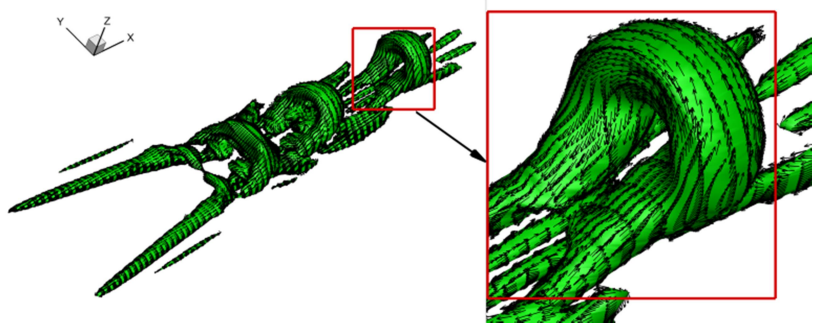

FIG. 13. The distribution of vortex vectors on the vortex cores which are illustrated by the iso-surface of $|\bar{R}| = 0.075$

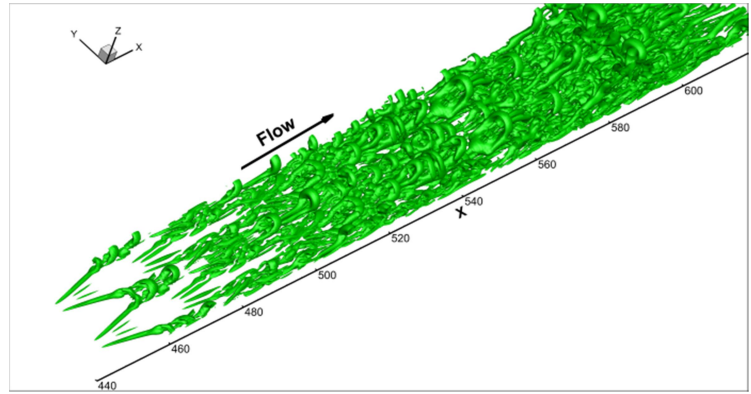

FIG. 14.The process of vortical structure "buildup" in laminar-turbulence flow transition visualized by the iso-surface of the magnitude of vortex vector $|\bar{R}| = 0.075$.

From the applications of vorticity and vortex vector on the vortices in the late boundary layer transition, it is found that vorticity line or tube can't represent vortex and vorticity can't well describe the rotation of fluid and the vortex structure of flow transition. However, the newly defined vortex vector can fully represent the fluid rotation and can fully describe the vortices, even if they are as complex as the vortices in the turbulence. The vortex can be visualized by the vortex line, vortex tube, or the iso-surface of the magnitude of the vortex vector. Different from the traditional vortex identification methods, vortex vector can not only visualize the structures of vortex cores but also show the rotation direction and the accurate strength of the vortex.

## C. Vortical structures after micro vortex generator (MVG)

For most vortex identification methods, there are problems when they are used in the compressible flow and modification is required[10,28]. However, from the mathematical definition of our new vortex vector, it can be seen that the derivation is based on the analysis of kinematics of fluid element and it is not related to the compressibility of the flow. Therefore, the vortex vector is only depended on the field of velocity and it can be used in any type of flow field. To show the ability in describing the complex vortical structures in the compressible flow, a MVG (Micro Vortex Generator) case with an inflow $Ma = 2.5$ is used as the third example.

MVG is a passive control device aiming to alleviate flow separation. The previous study[29] shows that MVG can produce strong vortex rings and it is the ring-like vortical structures which could reduce the flow separation. The computational domain is the same as given in Ref. **29** and the flow field is simulated by LES method in which the 5th order bandwidth-optimized WENO scheme is used.

The complex vortical structures near the MVG are visualized by the iso-surface of the magnitude of vortex vector $|\vec{R}| = 1$, see Fig. 15 in which the global view (Fig. 15(a)) and local view (Fig. 15(b)) of the vortical structures are presented. As we can see, vortex vector defined in this paper can clearly capture the large ring-like vortex structures and the small vortices at the same time even in such a complex case with interaction of the shock wave and turbulent boundary layers. The structures of vortex lines (blue lines in Fig. 15(b)) are also consistent with the cores of vortices. This reveals that the newly defined vortex vector can well describe the vortical structures in the compressible turbulence.

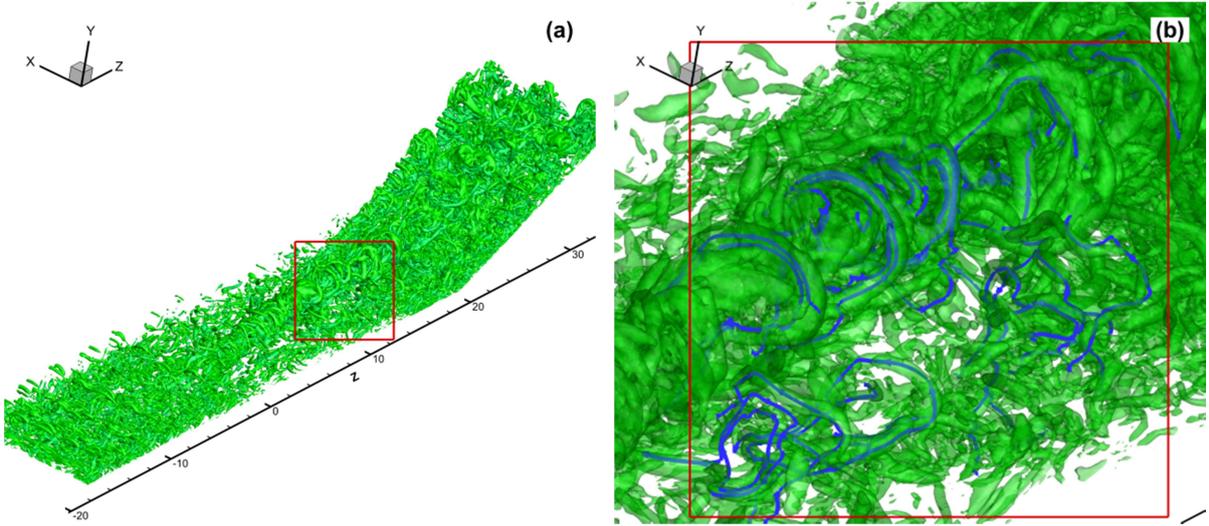

FIG. 15. Vortical structures of MVG case illustrated by the iso-surface of the magnitude of vortex vector $|\vec{R}| = 1$: (a)global view of vortical structures ; (b)local view of vortical structures with vortex lines created though points in several cores of vortices.

## V. Conclusions

Based on investigating the kinematics of fluid element in the 2D/3D flows, a definition of a new vector quantity called vortex vector is given in this paper to describe vortex. Three principles are followed during the definition of vortex vector: 1.The definition must be a local quantity as tens of thousands of vortices exist in turbulence and global or group quantity is not viable to define them; 2.The definition must be Galilean invariant; 3.The definition must be unique. Different from other vortex identification methods which are all scalar criterions, the newly defined variable is a vector quantity with a unique, Galilean invariant direction and a unique magnitude.

The vortex vector represents the local fluid rotation. The direction of the vortex vector is defined as the local fluid rotation axis along which the velocity components in the plane orthogonal to the vortex vector have zero derivatives in the vortex vector direction or $\frac{\partial U}{\partial Z} = 0$ and $\frac{\partial V}{\partial Z} = 0$ when the axis is set as parallel to the Z axis. The rotation axis is unique and Galilean invariant. The magnitude of the vortex vector is defined as the rotational part of vorticity in the direction of the vortex vector, which is twice of the minimum angular velocity of fluid around the point among all azimuth in the plane perpendicular to vortex vector and is also unique. Based on the definition of vortex vector, vortex is defined as a connected flow region where the magnitude of the vortex vector at each point is larger than zero. According to the definition of the vortex vector and vortex, the concepts of vortex line and vortex tube are also proposed to demonstrate the vortex structures.

The definitions of vortex vector and vector are evaluated using 2D/3D DNS /LES databases and are compared with vorticity and other vortex identification methods, such as Q-criterion and Ω-criterion. The results show that, unlike vorticity lines or tubes, the vortex lines or tubes based on the newly defined vortex vector are highly consistent with the cores of vortex which can be accurately represented by the iso-surface of the magnitude of vortex vector. Vortex vector can fully describe the vortical structures in not only incompressible flow but also compressible flow since it is derived only based on the kinematics of fluid element, but not fluid dynamics. Different from most vortex identification methods which only have a scalar variable and use the iso-surface to demonstrate the vortex structure, vortex vector, as a vector variable, can not only represent the accurate rotational strength of vortex but also can show the precise rotational direction, which is the advantage that most of vortex identification methods do not have.

As a rigorous mathematical definition with direction and magnitude, the newly defined vortex vector and vortex can be used to quantitatively investigate the physics of the generation and sustenance of vertical structures in turbulence. In the further work, we will study the transport equation for vortex vector and investigate the mechanics of the vortex generation and evolution in turbulence.

## ACKNOWLEDGMENTS

The authors are grateful to Texas Advanced Computing Center (TACC) for providing computation hours. The work was funded by the Priority Academic Program Development of Jiangsu Higher Education Institutions. This research was also supported in part by MURI FA9559-16-1-0364. This work is accomplished by using Code DNSUTA which was released by Dr. Chaoqun Liu at University of Texas at Arlington in 2009.